\documentclass[preprint,showpacs,preprintnumbers,amsmath,amssymb]{revtex4}
\usepackage{graphicx}
\usepackage{bm}
\usepackage{dcolumn}
\begin{document}
\begin{abstract}
The structure of the spin interaction operator (SI) ( the
interaction that remains  after space variables are integrated
out) in the first order S-matrix element of the elastic scattering
of a Dirac particle in
 a general helicity-conserving  vector potential is
 investigated.It is shown that the conservation
 of helicity dictates  a specific form of the SI regardless of the
 explicit form of the vector potential. This SI closes the $SU(2)$
 algebra with other two operators in the spin space of the
 particle. The directions of the momentum transfer vector and the
 vector bisecting the scattering angle seem to define some sort of
 "intrinsic" axes at this order that act as some symmetry axes  for the whole spin
 dynamics . The conservation of
 helicity at this order can be formulated as the invariance of the
 component of the helicity of the particle along the bisector of the scattering angle
 in the transition.

\end{abstract}
\title{Conservation of Helicity and $SU(2)$ Symmetry in
 First Order Scattering }
\author{A.Albaid}
\author{M.S.Shikakhwa}
\email{moody@ju.edu.jo}%
\address{Department of Physics, University of Jordan\\
11942--Amman, Jordan}%

\maketitle


\maketitle
\section{Introduction}
It is well-known that the helicity of a Dirac particle is
conserved given that there is no electric field acting on the
particle \cite{sakurai}.  Indeed, the Heisenberg equation of
motion for the helicity operator $\mathbf{\Sigma}.\mathbf{\Pi}$
where $\mathbf{\Pi}=(\mathbf{p}-e\mathbf{A})$ is the mechanical
momentum of the particle reads ($\hbar=c=1$):
\begin{equation}\label{1}
[ \mathbf{\Sigma}.\mathbf{\Pi},H]=e\mathbf{\Sigma}\cdot\mathbf{E}
\end{equation}
Here, $H=H_0+H_{\mathrm{int}}$, with
\begin{equation}\label{}
\label{2}H_0=\boldsymbol{\alpha}\cdot\mathbf{p}+\beta m,
\end{equation}
and
\begin{equation}
\label{3}H_{\mathrm{int}}=eA_0-e\boldsymbol{\alpha}\cdot\mathbf{A},
\end{equation}
 $e$ is the charge of the particle, ${\alpha}_i=\beta\gamma_i$
and $\beta=\gamma_4$. The $\gamma$'s are the Dirac matrices:
$\{\gamma_\mu,\gamma_\nu\}=2g_{\mu\nu}$. Thus, in the absence of
an electric field $\mathbf{E}$ helicity is conserved. In physical
terms, conservation of helicity is described as the invariance of
the component  of the spin of the particle along its momentum. In
the perturbative expansion of a helicity-conserving theory,
helicity is conserved at each order of the perturbation series .
For example, in the first order S-matrix element of the elastic
scattering of a particle in some helicity-conserving vector
potential, the conservation of helicity manifests itself through
the fact that if the incident particle is in an eigen state of the
operator $\mathbf{\Sigma}.\mathbf{\hat{p}}_i$
($\mathbf{\hat{p}}_i\equiv\frac{\mathbf{p}_i}{|\mathbf{p}_i|}$)
then the interaction will map it onto an eigenstate of
$\mathbf{\Sigma}.\mathbf{\hat{p}}_f$ with the same eigenvalue
\cite{sakurai} ( $\mathbf{p}_i$ and $\mathbf{p}_f$ are the
incident and outgoing momenta, respectively). We can view this as
if the interaction itself rotates the spin of the particle with
the same angle $\theta$ with which it deviates the incident
momentum so that the spin projection along the momentum is
conserved in the transition. This work focuses on the conservation
of helicity at this order and attempts to investigate the
structure and properties of the spin interaction operator (SI) -
the interaction operator that remains in the first order S-matrix
of the scattering of a single Dirac particle after integrating out
the space degrees of freedom- for a general helicity-conserving
vector potentials. It will be shown that conservation of helicity
at this order demands that this SI has a specific  form regardless
of the explicit form of the vector potential.As a concrete
example, the Aharonov-Bohm (AB) potential \cite{AB} treated in an
earlier work \cite{helicity} will be considered .
\section{ Effective Spin Interaction}
Consider a Dirac particle in a given magnetic field whose vector
potential is the static vector function $\mathbf{A\mathbf(x)}$ and
such that there is no scalar potential. The first order S-matrix
element for the elastic scattering of a particle in this potential
is :
\begin{equation}
\label{5}S_{fi}^{(1)}= i\int {d^4 x\,\bar \psi _f \left( x
\right)\left( e\mathbf{\gamma}\cdot\mathbf{A} \right)\psi _i
\left( x \right)}.
\end{equation}
Carrying out the time integral,we get this as
\begin{equation}
\label{5b}S_{fi}^{(1)}=- 2\pi e|N|^{2} \delta (E_f-E_i)
u^{\dagger} _f \left( p_f,s_f \right)\left(\int {d^3
x\,e^{i\left(\mathbf{p}_f-\mathbf{p}_i\right)\cdot\mathbf{x}}\left(
\mathbf{\alpha}\cdot \mathbf{A} \right)}\right)u _i \left(p_i,s_i
\right).
\end{equation}
which can be casted in the form
\begin{equation}
\label{5c}S_{fi}^{(1)}=- 2\pi e|N|^{2} \delta (E_f-E_i)
u^{\dagger} _f \left( p_f,s_f \right)\left( \mathbf{\alpha}\cdot
\mathbf{A}(\mathbf{q}) \right)u _i \left(p_i,s_i \right).
\end{equation}
where $\mathbf{A}(\mathbf{q})$ is the Fourier transform of the
vector potential with respect to the momentum transfer vector
$\mathbf{q}=\mathbf{p}_f-\mathbf{p}_i$ and $N$ is a normalization
constant . Recalling that $\alpha_i=\gamma_5\Sigma_i$, where
$\Sigma_i=\frac{i}{2}[\gamma_i,\gamma_j]\qquad ,\quad (i,j=1..3)$,
and $i\gamma_5=\gamma_1\gamma_2\gamma_3\gamma_4$, we write the
matrix element as:
\begin{equation}
 \label{5}S_{fi}^{(1)}=- 2\pi e|N|^{2}|\mathbf{A}(\mathbf{q})| \delta
(E_f-E_i) u^{\dagger} _f \left( p_f,s_f \right)\left(
\gamma_5\mathbf{\Sigma}.\mathbf{\hat{a}} \right)u _i \left(p_i,s_i
\right).
\end{equation}
where we have introduced the unit vector
$\mathbf{\hat{a}}=\frac{\mathbf{A}(\mathbf{q})}{|\mathbf{A}(\mathbf{q})|}$.
The operator $K\equiv \gamma_5\mathbf{\Sigma}.\mathbf{\hat{a}}$ is
what we denote with the spin
 interaction operator (SI) as it is the operator that induces transition in the spin space of
 the particle. Since we are considering the scattering of a
single particle, then it is always possible to find a plane that
contains both the incident and outgoing momenta. We choose the
coordinates so that the positive $x$-axis is defined by the
direction of the incident momentum $\mathbf{p}_i$, and take the
$y$-axis to be normal to it in this plane. In this coordinate
system, we have $\mathbf{p}_i=(p,0,0)$ and
$\mathbf{p}_f=(p\cos\theta,p\sin\theta,0)$.Obviously, in this
coordinate system $\mathbf{q}$ is planar as well. Now, we impose
conservation of helicity on the matrix element and deduce the
consequences of this on $K$. Let us use the Dirac notation and
denote with $|\mathbf{p_i};\pm>$ the eigenstates of
$\mathbf{\Sigma}.\mathbf{\hat{p}}_i$ with eigenvalues $\pm 1$.
Similarly, $|\mathbf{p_f};\pm>$ are the eigenstates of
 $\mathbf{\Sigma}.\mathbf{\hat{p}}_f$.Now, conservation of helicity in
 the matrix element, Eq.(\ref{5}) means that if the incident
 particle is in an eigenstate $|\mathbf{p_i};\pm>$ of
 $\mathbf{\Sigma}.\mathbf{\hat{p}}_i$, then the state
 $K|\mathbf{p_i};\pm>$ is an eigenstate of
 $\mathbf{\Sigma}.\mathbf{\hat{p}}_f$ with the same eigenvalue,i.e
 \begin{equation}\label{8}
 \mathbf{\Sigma}.\mathbf{\hat{p}}_f K|\mathbf{p_i};\pm>=\pm K|\mathbf{p_i};\pm>
\end{equation}
Since we can write $|\mathbf{p_i};\pm>=\pm
\mathbf{\Sigma}.\mathbf{\hat{p}}_i |\mathbf{p_i};\pm>$, we have
Eq.(\ref{8}) as
\begin{equation}\label{}
 \pm \mathbf{\Sigma}.\mathbf{\hat{p}}_f K\mathbf{\Sigma}.\mathbf{\hat{p}}_i|\mathbf{p_i};\pm>=\pm K|\mathbf{p_i};\pm>
\end{equation}
from which we deduce the following relation among the operators in
the helicity space
\begin{equation}\label{10}
  \mathbf{\Sigma}.\mathbf{\hat{p}}_f K\mathbf{\Sigma}.\mathbf{\hat{p}}_i=K.
\end{equation}
Now, substituting the explicit forms of the operators in the above
equation
($\mathbf{\Sigma}.\mathbf{\hat{p}}_f=\Sigma_1\cos\theta+\Sigma_2\sin\theta,
\quad\mathbf{\Sigma}.\mathbf{\hat{p}}_i=\Sigma_1$ and
$K=\gamma_5\mathbf{\Sigma}.\mathbf{\hat{a}}$, $\theta$ being the
scattering angle) we get the following conditions on the
components of the unit vector $\mathbf{\hat{a}}$:
\begin{eqnarray}
a_1\cos\theta+a_2\sin\theta&=&a_1\nonumber\\
a_1\sin\theta-a_2\cos\theta&=&a_2\nonumber\\
a_3&=&0.
\end{eqnarray}
The solution of the above set of equations is
$\mathbf{\hat{a}}=\pm\mathbf{\hat{k}}$, where
$\mathbf{\hat{k}}=\frac{\mathbf{p_f}+\mathbf{p_i}}
{|\mathbf{p_f}+\mathbf{p_i}|}=(\cos\frac{\theta}{2},\sin\frac{\theta}{2},0)$
 is a vector bisecting the scattering angle $\theta$ ( see Fig.1 ) and normal to the
momentum transfer vector $\mathbf{q}$. Evidently, the third
component of $\mathbf{\hat{a}}$ vanishes as expected with our
current choice of axes. Thus, adopting the positive sign of the
solutions, we have now the SI as :
\begin{equation}\label{10b}
K=\gamma_5\mathbf{\Sigma.\hat {k}}.
\end{equation}
\begin{figure}[htbp]
\includegraphics[width=8cm]{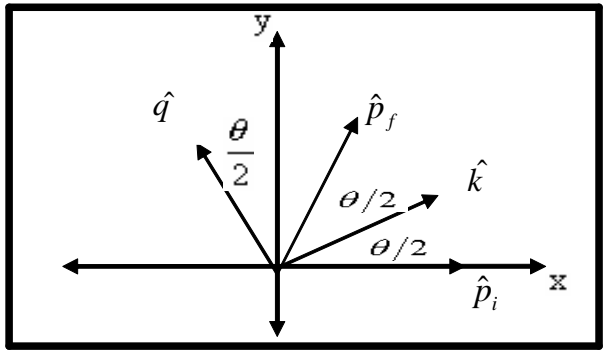}
\caption{\label{fig1} Scattering diagram in the xy-plane.}
\end{figure}
The above is, therefore, the specific form of the spin interaction
in the first order S-matrix element of the elastic scattering of a
single particle in any helicity-conserving vector potential
regardless of its explicit form. Only applicability of the Born
approximation is required. The above Hermitian SI operator $K$
(more precisely $\gamma_5K$) closes the $SU(2)$ algebra with other
two operators in the spin space of the particle:
\begin{eqnarray}\label{11}
  [\mathbf{\Sigma}.\mathbf{\hat{k}},\mathbf{\Sigma}.\mathbf{\hat{q}}] &=& 2i\Sigma_3 \nonumber\\
\left[\Sigma_3,\mathbf{\Sigma}.\mathbf{\hat{k}}\right] &=& 2i\mathbf{\Sigma.\hat{q}}\ \\
  \left[\mathbf{\Sigma}.\mathbf{\hat{q}},\Sigma_3\right] &=&
  2i\mathbf{\Sigma.\hat{k}},\nonumber.
\end{eqnarray}
The following algebra can also be easily verified:
\begin{equation}\label{12}
\left\{\mathbf{\Sigma}.\mathbf{\hat{k}},\mathbf{\Sigma}.\mathbf{\hat{q}}\right\}=
\left\{\Sigma_3,\mathbf{\Sigma}.\mathbf{\hat{k}}\right\}=\left\{\mathbf{\Sigma}.\mathbf{\hat{q}},\Sigma_3\right\}=0
\end{equation}
Thus, the consequences:
\begin{equation}\label{13}
(\mathbf{\Sigma.\hat{k}})^2=(\mathbf{\Sigma.\hat{q}})^2=(\Sigma_3)^2=I.
\end{equation}
The above results are suggestive of an interesting picture: Since
the vectors $\mathbf{\hat{q}}$ and $\mathbf{\hat{k}}$ are
"intrinsic", in that they are defined by the dynamics of the
system, then it seems that a helicity-conserving  system makes a
natural choice of the representation of the generators of the
$SU(2)$ in the spin space that goes along with its dynamics. The
SI operator being one of these. Actually, the whole of the spin
dynamics can be formulated in terms of the new "intrinsic" axes
and the corresponding spin operators ( see Fig.2). To see this,
let us express the helicity operators in terms of the new
generators $\mathbf{\Sigma.\hat{k}}$ and $\mathbf{\Sigma.\hat{q}}$
as
\begin{eqnarray}
  \label{15}
\mathbf{\Sigma.\hat{p_i}}&=&\cos\frac{\theta}{2}\mathbf{\Sigma.\hat{k}}-\sin\frac{\theta}{2}\mathbf{\Sigma.\hat{q}}\nonumber\\
\mathbf{\Sigma.\hat{p_f}}&=&\cos\frac{\theta}{2}\mathbf{\Sigma.\hat{k}}+\sin\frac{\theta}{2}\mathbf{\Sigma.\hat{q}}
\end{eqnarray}
\begin{figure}[htbp]
\includegraphics[width=8cm]{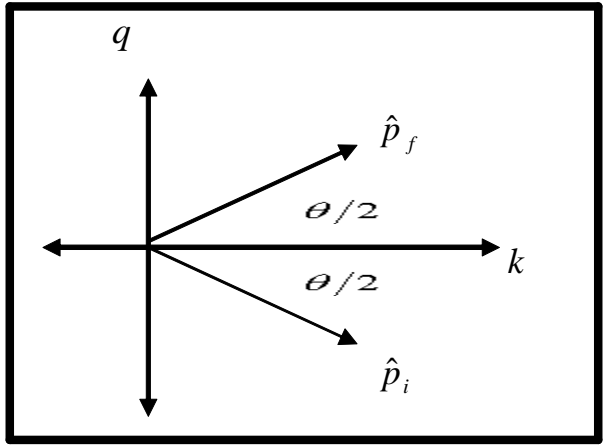}
\caption{\label{fig2} Scattering diagram in the $k-q$ plane.}
\end{figure}
Note the symmetry involved in these two expressions which is a
reflection of the symmetry involved in Fig.2. With the above
expressions in hand, it is easy to derive the following relations
 using the algebra , Eqs.(\ref{12});
\begin{eqnarray}\label{17}
K\mathbf{\Sigma.\hat{p_i}}K&=& \mathbf{\Sigma.\hat{p_f}} \nonumber\\
K\mathbf{\Sigma.\hat{p_f}}K&=& \mathbf{\Sigma.\hat{p_i}},
\end{eqnarray}
from which we can easily rederive Eq.(\ref{10});
\begin{equation}\label{18}
\mathbf{\Sigma.p_f}K\mathbf{\Sigma.p_i}=K.\nonumber
\end{equation}
Eq.(\ref{17}) suggests that the incident and outgoing particles'
helicity operators are related through a reflection across the
$\mathbf{\hat{k}}-$axis induced by the SI, $K$, and conservation
of helicity in the transition can be expressed as the invariance
of the components of the initial and final helicity states  along
the $\mathbf{\hat{k}}$-axis. Indeed, it is possible to express the
eigenstates of  $\mathbf{\Sigma.\hat{p_i}}$ and
$\mathbf{\Sigma.\hat{p_f}}$ in terms of the eigenstates of
$\mathbf{\Sigma .\hat{k}}$, viz.
\begin{eqnarray}\label{22}
  |\hat{p}_i,\pm\rangle &=&  \cos\frac{\theta}{4}|\hat{k},\pm\rangle\mp\sin\frac{\theta}{4}|\hat{k},\mp\rangle\nonumber\\
 |\hat{p}_f,\pm\rangle &=&
 \cos\frac{\theta}{4}|\hat{k},\pm\rangle\pm\sin\frac{\theta}{4}|\hat{k},\mp\rangle,
  \end{eqnarray}
  where use has been made of
  $\mathbf{\Sigma.\hat{q}}|\hat{k},\pm\rangle=|\hat{k},\mp\rangle$,
  which can be proven easily using the algebra,Eqs.(\ref{11}) and (\ref{12}). The
  following result then, follows immediately:
  \begin{equation}
\langle\hat{k},\pm|\hat{p}_i,\pm\rangle=\langle\hat{k},\pm|\hat{p}_f,\pm\rangle=\cos\frac{\theta}{4}.
\end{equation}
The component of the spin of the particle along the
$\mathbf{\hat{k}}$-axis is indeed invariant in the transition.

\section{Aharonov-Bohm Vector Potential}
In this section we consider as a concrete example of a vector
potential the Aharonov-Bohm (AB) vector potential \cite{AB}. We
will show that the form of the SI, $K$, given by Eq.(\ref{10b})
follows for this potential. The (AB) potential conserves helicity
in view of Eq.(\ref{1})as it does not generate an electric field.
The conservation of helicity in the first order S-matrix was
verified using explicit wave functions  in \cite{vera}.
$\mathbf{A}(\mathbf{x})$ for this potential is planar and reads
\begin{equation}
\label{7}\mathbf{A}=\frac{\Phi}{2\pi\rho}\hat{\epsilon}_\varphi,
\end{equation}
where $\rho=\sqrt{x^2+y^2}$, $\hat{\epsilon}_\varphi$ is the unit
vector in the $\varphi$-direction, and $\Phi$ is the flux through
the AB tube. Since the vector potential is planar , the dynamics
is essentially planar by default in this case, and we can always
take
$\mathbf{p_i}=(p,0,0)\;,\;\mathbf{p_f}=(p\cos\theta,p\sin\theta,0)$,
$\theta $ being the scattering angle. The first order S-matrix
element,Eq.(\ref{5b}), upon plugging the above expression for
$\mathbf{A}(\mathbf{x})$ then gives :
\begin{equation}\label{7}
S^{(1)}_{fi}=-4\pi^2\Delta |N|^{2} \delta
(E_f-E_i)\frac{1}{q}<\mathbf{p_f};s_f|\frac{\alpha_1 q_2-\alpha_2
q_1}{q}|\mathbf{p_i};s_i>
\end{equation}
Here, $\Delta=-e\Phi/2\pi$ (in perturbative calculations
$0<\Delta<1$).Thus, comparing with Eq.(\ref{5c}), we identify
$\frac{1}{q}$ with $|\mathbf{A}(\mathbf{q})|$ and $(\frac{\alpha_1
q_2-\alpha_2 q_1}{q})$ with $K$. It is easily checked that the
above $K$ indeed has the  specific form given by Eq.(\ref{10b});
\begin{equation}\label{9}
K=(\frac{\alpha_1 q_2-\alpha_2
q_1}{q})=\gamma_5\mathbf{\Sigma}.\mathbf{\hat{k}}\nonumber
\end{equation}
\section{Conclusions} The dynamics of the elastic scattering of
a Dirac particle in a vector potential ( magnetic field) in the
first order Born approximation can always be reduced to a planar
one by choosing a specific set of the coordinate axes.  Thus, we
have shown that the spin interaction (SI) ( the interaction that
remains after integrating out the space degrees of freedom) in the
first order scattering matrix for a  general helicity-conserving
vector potential assumes the specific form
$K=\gamma_5\mathbf{\Sigma}.\hat{k}$, where
$(\mathbf{\hat{k}}=\frac{\mathbf{p_f}+\mathbf{p_i}}
{|\mathbf{p_f}+\mathbf{p_i}|}=(\cos\frac{\theta}{2},\sin\frac{\theta}{2},0)$
is a unit vector in the scattering plane that bisects the
scattering angle $\theta$. The operator $\mathbf{\Sigma}.\hat{k}$
 along  with the two
operators
$\mathbf{\Sigma.\hat{q}}\,(\hat{q}=\frac{\mathbf{q}}{q},\,\mathbf{q}=\mathbf{p_f-p_i})$
and $\Sigma_3$  were shown to close the $SU(2)$. This means that
$\mathbf{\Sigma.\hat{k}}$ and $\mathbf{\Sigma.\hat{q}}$ can be
identified - in addition to $\Sigma_3$- as furnishing a
representation of the generators of this group in the spin space
of the particle. What is interesting with this representation is
the fact that it is " intrinsic' in that the vectors $\mathbf{k}$
and $\mathbf{q}$ are defined by the dynamics of the system itself.
It was shown that expressing the helicity operators of the initial
and final states in terms of $\mathbf{\Sigma.\hat{k}}$ and
$\mathbf{\Sigma.\hat{q}}$, the conservation of helicity in the
transition can be viewed as the invariance of the
$\mathbf{k}$-component of the helicity of the particle in the
transition. As a concrete example , the AB potential was
considered and the specific form of the SI was shown to emerge.

\begin{acknowledgments}
We are indebted to Professor H.J.Weber and Dr. K. Bodoor for
reading the manuscript and helpful suggestions and discussions.
\end{acknowledgments}
\bibliography{helicity}

\end{document}